\documentclass{aa}
\usepackage{graphicx}
\usepackage{epsfig}
\def\deg{\ifmmode^{\circ}\;\else$^{\circ}\;$\fi} 
\def\lsim{\,\lower2truept\hbox{${<\atop\hbox{\raise4truept\hbox{$\sim$}}}$}\,}
\def\gsim{\,\lower2truept\hbox{${> \atop\hbox{\raise4truept\hbox{$\sim$}}}$}\,}
\begin{document}
   \title{Polarization Properties of Extragalactic Radio Sources
   \\
   and Their Contribution to Microwave Polarization Fluctuations}


   \author{D. Mesa
          \inst{1}
          \and
          C. Baccigalupi\inst{2,3}
          \and
          G. De Zotti\inst{4}
       \and
          L. Gregorini\inst{5,6}
            \and
          K.-H. Mack\inst{8,5,7}
          \and
          M. Vigotti\inst{5}
           \and
          U. Klein\inst{7}
          }
   \offprints{G. De Zotti}

   \institute{Dipartimento di Astronomia, Universit\`a di Padova, Vicolo
dell'Osservatorio 2, I-35122 Padova, Italy\\
e-mail: mesa@mostro.pd.astro.it
   \and
SISSA, International School for Advanced Studies, Via Beirut 2-4,
I-34014 Trieste, Italy\\
e-mail: bacci@sissa.it \and Lawrence Berkeley National Laboratory,
1 Cyclotron Road, Mailstop 50-205, Berkeley, CA 94720, USA \and
Osservatorio Astronomico di Padova,
INAF, Vicolo dell'Osservatorio 5, I-35122 Padova, Italy  \\
email: dezotti@pd.astro.it \and IRA/CNR, Via Gobetti 101, I-40129
Bologna, Italy \and Dipartimento di Fisica, Universit\`a di
Bologna, Via Irnerio 46, I-40126 Bologna, Italy \and
Radioastronomisches Institut der Universit\"at Bonn, Auf dem
H\"ugel 71, D-53121 Bonn, Germany \and ASTRON/NFRA, Postbus 2,
NL-7990 AA Dwingeloo, The Netherlands \\
e-mail: mack@astron.nl
             }

\date{Received 05-07-02; accepted 23-09-02}

\abstract{We investigate the statistical properties of the
polarized emission of extragalactic radio sources and estimate
their contribution to the power spectrum of polarization
fluctuations in the microwave region. The basic ingredients of our
analysis are the NVSS polarization data, the multifrequency study
of polarization properties of the B3-VLA sample (Mack et al. 2002)
which has allowed us to quantify Faraday depolarization effects,
and the 15 GHz survey by Taylor et al. (2001), which has provided
strong constraints on the high-frequency spectral indices of
sources. The polarization degree of both steep- and flat-spectrum
at 1.4 GHz is found to be anti-correlated with the flux density.
The median polarization degree at 1.4 GHz of both steep- and
flat-spectrum sources brighter than $S(1.4\,\hbox{GHz})=80\,$mJy
is $\simeq 2.2\%$. The data by Mack et al. (2002) indicate a
substantial mean Faraday depolarization at 1.4 GHz for steep
spectrum sources, while the depolarization is undetermined for
most flat/inverted-spectrum sources. Exploiting this complex of
information we have estimated the power spectrum of polarization
fluctuations due to extragalactic radio sources at microwave
frequencies.
We confirm that extragalactic sources are expected to be the main
contaminant of Cosmic Microwave Background (CMB) polarization maps
on small angular scales. At frequencies $< 30\,$GHz the amplitude
of their power spectrum is expected to be comparable to that of
the $E$-mode of the CMB. At higher frequencies, however, the CMB
dominates.


   }
\titlerunning{Polarization Properties of Extragalactic Radio
Sources}

   \maketitle
%

\begin{figure*}
  \centering
 \includegraphics[width=7in,height=3.5in]{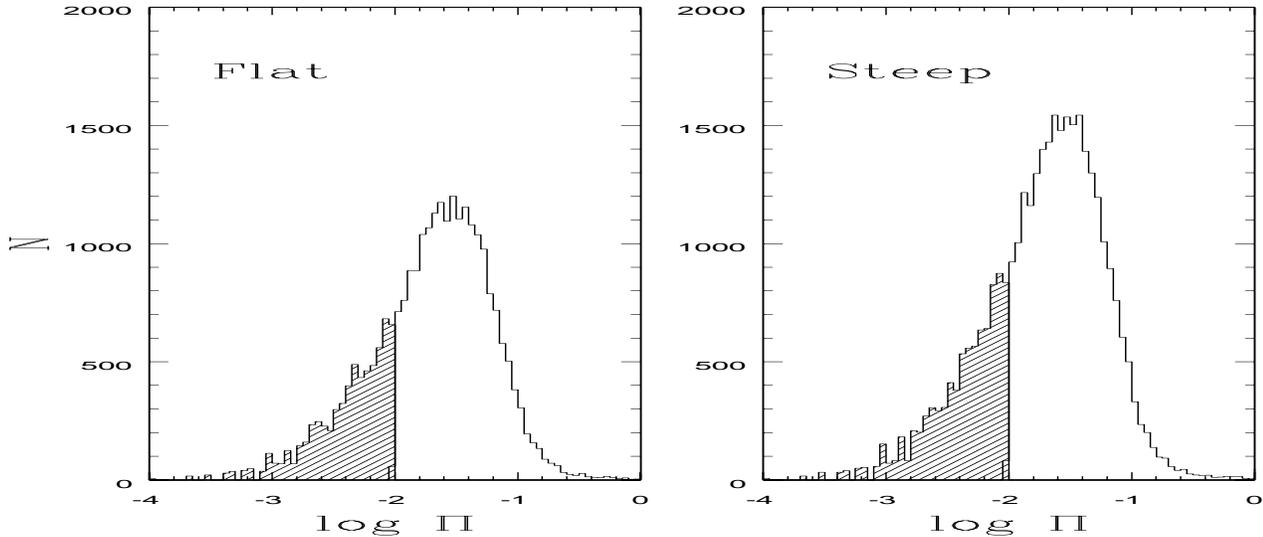}
      \caption{Polarization degree distributions of flat/inverted- (left) and
      steep-spectrum  (right) NVSS sources brighter than 80 mJy. The
      hatched areas correspond to values seriously contaminated by
      instrumental polarization (see text). The median
      polarization degree is 2.2\%, for both populations.
              }
         \label{poldistrflatsteep}
   \end{figure*}
%
%
  \begin{figure*}
   \centering
\includegraphics[width=7in,height=6in]{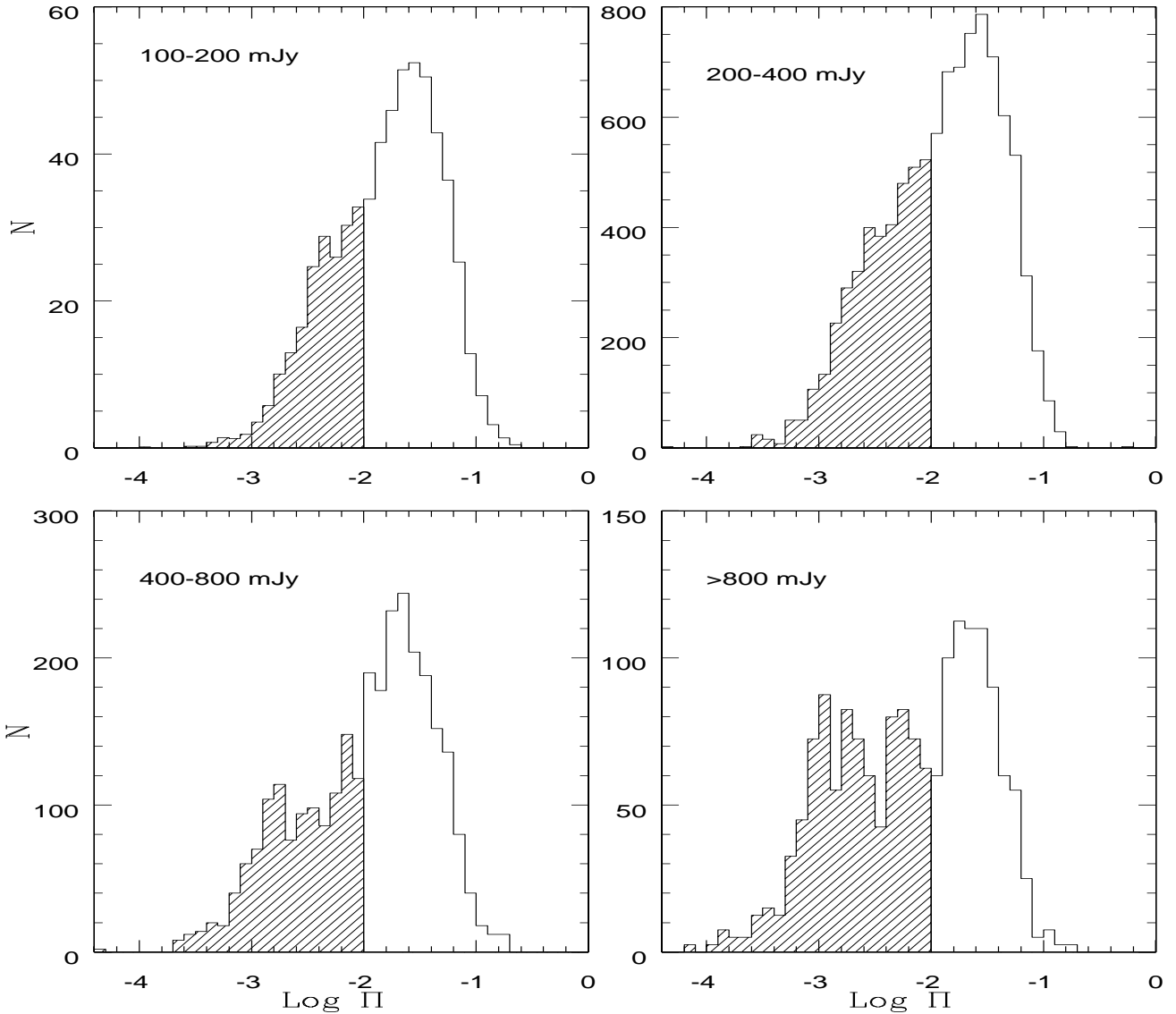}
\caption{Polarization degree distribution of steep-spectrum
      NVSS sources for several flux density intervals: 100--200 mJy
      (8032 sources; median value: 1.82\%), 200--400 mJy (3700 sources;
      median value: 1.45\%), 400--800 mJy (1438 sources; median value: 1.37\%),
      $>800\,$mJy (660 sources; median value $<1\%$, formal median value: 0.74\%).
      The hatched areas have the same meaning as in the previous figure.
              }
         \label{poldegvsfluxs}
   \end{figure*}
%
%
  \begin{figure*}
   \centering
\includegraphics[width=7in,height=6in]{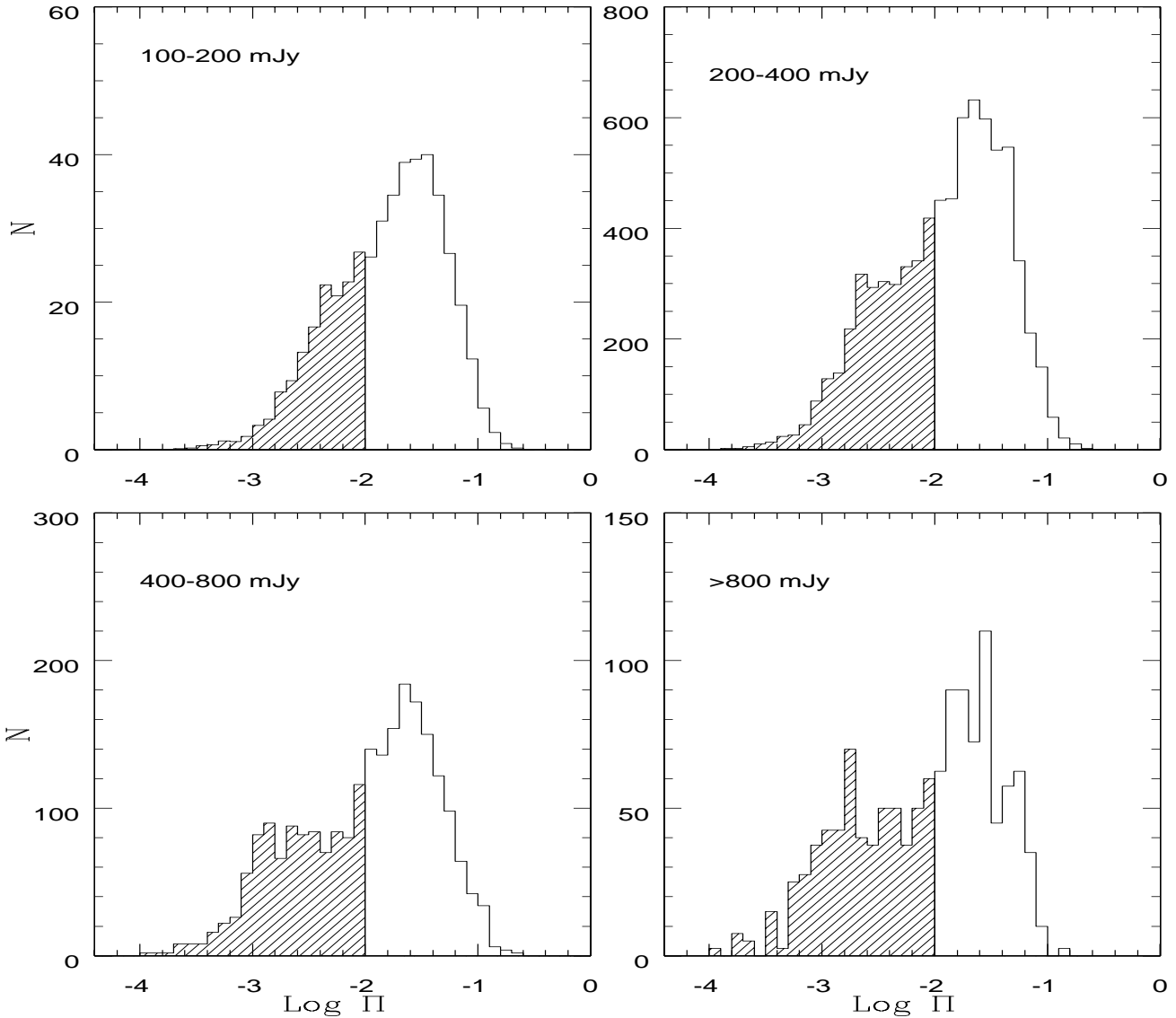}
      \caption{Polarization degree distribution of flat/inverted-spectrum
      NVSS sources for several flux density intervals: 100-200 mJy
      (6198 sources; median value: 1.84\%), 200-400 mJy (2859 sources; median value: 1.50\%),
      400--800 mJy (1150 sources; median value: 1.32\%), $>800\,$mJy
      (496 sources; median value: 1.05\%). The hatched areas correspond to values seriously
      contaminated by instrumental polarization (see text).
              }
         \label{poldegvsfluxf}
   \end{figure*}

  \begin{figure*}
   \centering
\includegraphics[width=7in,height=3in]{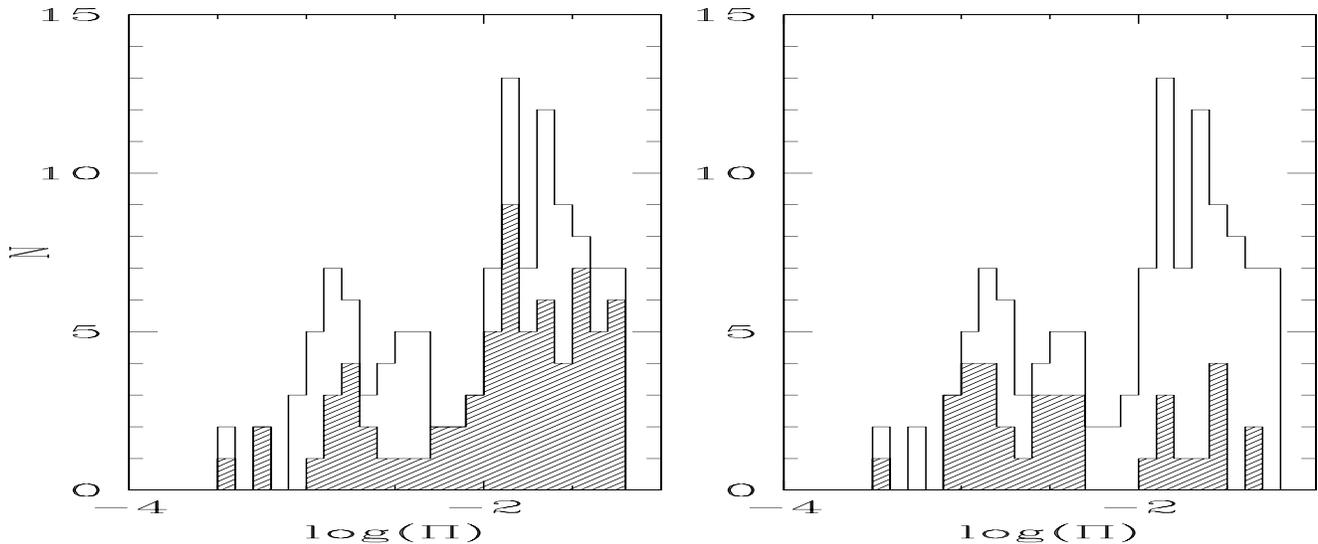}
      \caption{Polarization degree distribution of flat/inverted-spectrum
      NVSS sources with $S(1.4\hbox{GHz}) \ge 500\,$mJy optically identified
      by Snellen et al. (2002). The hatched area in the left panel
      shows the distribution of stellar sources, the one in the
      right panel, the distribution of extended sources.
                   }
         \label{distrpolsnellen}
   \end{figure*}

  \begin{figure}[h]
   \centering
\includegraphics[width=3in,height=4in]{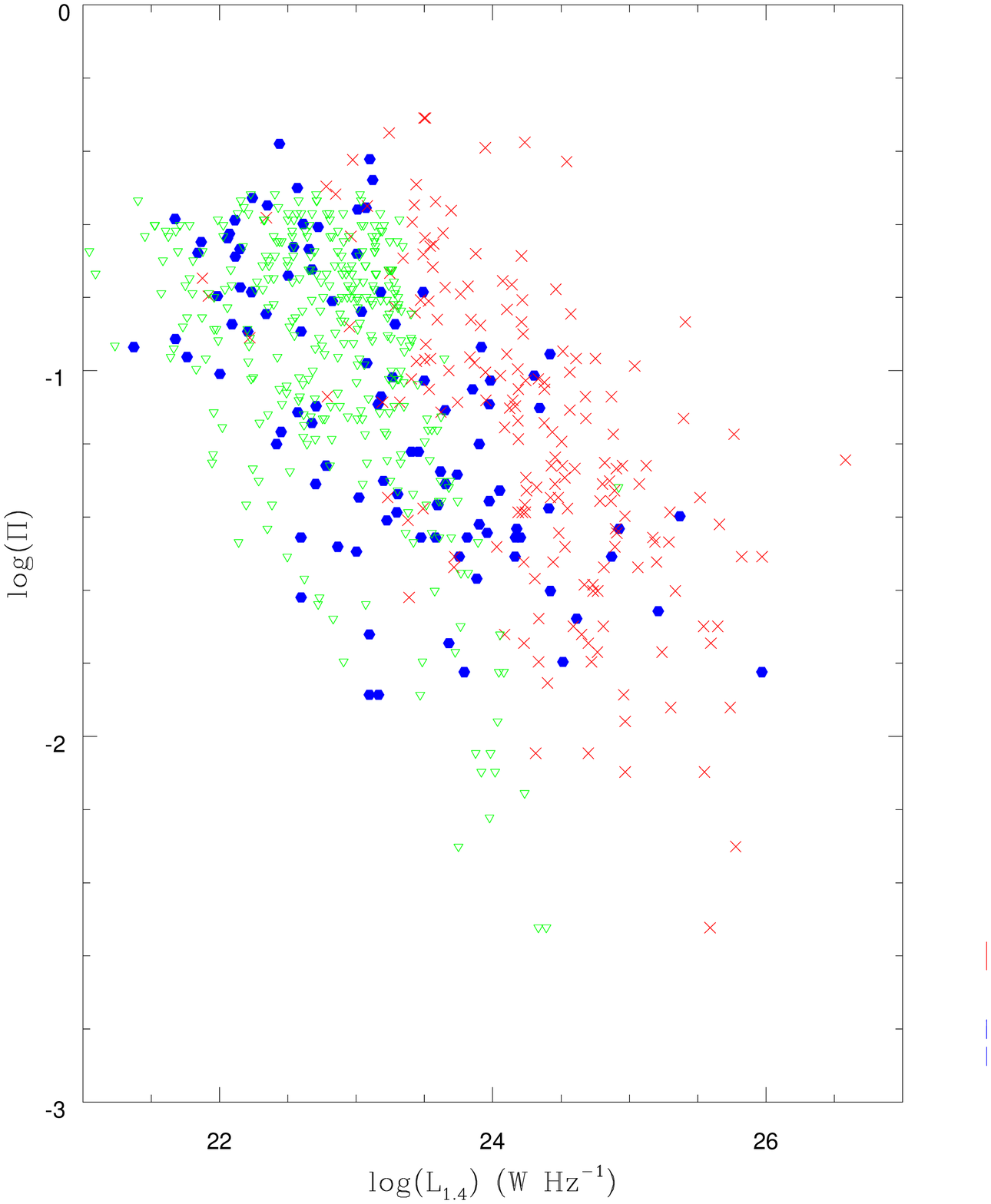}
      \caption{Observed polarization degree versus radio luminosity for NVSS galaxies
      in the sample by Sadler et al. (2002). Crosses are AGNs, filled hexagons
      are star-forming galaxies. Sources whose polarization is either not measured
      to better than $3\sigma$ or is $< 1\%$ (and therefore substantially
      contaminated by instrumental polarization) have been attributed an upper
      limit equal to the maximum between $3\sigma$ and 1\%. These upper limits are
      shown as triangles. An Einstein-de Sitter universe with
      $H_0=50$ has been adopted.
                   }
         \label{polvslumEdS}
   \end{figure}

\section{Introduction}

Polarization measurements provide crucial information on the
physics of radio sources. At high enough frequencies for Faraday
rotation to be negligible we can reliably assume that the magnetic
field direction lies perpendicular to the observed polarization
position angle. On the other hand, determinations of the Faraday
rotation measures (RMs) are informative on the magneto-ionic
properties of the medium embedding the emitting region or along
its line-of sight.

Another very important use of polarization measurements of large
samples of radio sources is to quantify the contamination by these
sources of polarization maps of the Cosmic Microwave Background
(CMB). The astonishing advances in our understanding of the basic
properties of the Universe and in precision determinations of its
fundamental parameters made possible by the recent accurate
measurements of acoustic peaks of the Cosmic Microwave Background
(CMB) anisotropy power spectrum by the TOCO (Miller et al. 1999),
BOOMERanG (de Bernardis et al. 2002), MAXIMA (Lee et al. 2001),
DASI (Halverson et al. 2002), and CBI (Pearson et al. 2002)
experiments, have put further impetus in experimental efforts to
exploit the extraordinary wealth of cosmological information
carried by the CMB.

A Gaussian CMB fluctuation field is fully characterized by four
power spectra, $C_\ell^{TT}$, $C_\ell^{EE}$, $C_\ell^{BB}$,
$C_\ell^{TE}$, where $T$ stands for ``temperature'', while $E$ and
$B$ are rotationally invariant fields, which are linear, but
non-local, combinations of the Stokes parameters Q and U (Seljak
1997; Kamionkowski et al. 1997; Zaldarriaga \& Seljak 1997). It is
then clear that polarization measurements are crucial to fully
exploit the CMB information content. In particular, detection of
CMB polarization is critical for tests of Planck-scale physics (Hu
\& White 1997; Kamionkowski \& Kosowski 1999; Hu \& Dodelson
2002). This has motivated experimental efforts by many groups (see
Staggs et al. 1999; Cecchini et al. 2002; De Zotti 2002). Although
these measurements are very challenging because of the weakness of
the expected signal (at, or below, several $\mu$K level), recent
upper limits (Hedman et al. 2001; Keating et al. 2001) have
already got close to it, and a detection may be achieved in the
next few years.

It is not yet clear, however, whether the ultimate limit to our
ability of measuring the CMB polarization will be set by detector
sensitivity or by foregrounds, because the latter are still very
poorly understood. On small angular scales, foreground intensity
fluctuations at cm and mm wavelengths are dominated by
extragalactic radio sources (Toffolatti et al. 1998, 1999), which
are significantly polarized and are therefore expected to dominate
also foreground polarization fluctuations up to at least $\sim
100\,$GHz. Preliminary investigations have been carried out by
Sahzin \& Kor\"olev (1985) and De Zotti et al. (1999). These works
assumed a constant mean polarization degree for all classes of
radio sources (estimated from rather small samples), a Poisson
space distribution, and adopted mean values of the spectral
indices to extrapolate to high frequencies.

The NRAO VLA Sky Survey (NVSS, Condon et al. 1998), covering
$\simeq 82\%$ of the sky to a flux density limit of $\simeq
2.5\,$mJy at 1.4 GHz and containing data on Stokes $I$, $U$, and
$Q$ parameters for almost $2\times 10^6$ sources, has provided an
extensive data base for a statistical investigation of the
polarization properties of extragalactic sources. Important
complementary information comes from the multifrequency study of
over 100 sources drawn from the B3-VLA sample (Mack et al. 2002),
which allows us to get insight into the effect of Faraday
depolarization, which strongly affects polarization measurements
at 1.4 GHz. Our analysis is presented in Sect.$\,2$. The data by
Condon et al. (1998) and Mack et al. (2002) also allow us to
considerably improve on the available estimates of polarization
fluctuations due to extragalactic sources in the microwave region.
The new analysis, presented in Sect.$\,3$, exploits the real space
distribution of NVSS sources, which covers the flux density range
relevant for CMB experiments, the true 1.4--5.85$\,$GHz spectral
index distribution of sources, obtained combining NVSS and GB6
(Gregory et al. 1996) data, the polarization degree distribution
at 1.4 GHz, and the correction for Faraday depolarization, based
on the data by Mack et al. (2002). The new constraints on the high
frequency spectral indices set by the 15 GHz survey by Taylor et
al. (2001) are also taken into account. Our main conclusions are
summarized and discussed in Sect.$\,4$.

%
%

 %

\section{Polarization Properties of Extragalactic Radio Sources}

The 1.4--4.85$\,$GHz spectral indices $\alpha\ (S_\nu \propto
\nu^{-\alpha})$ of a complete sub-set of NVSS sources were
obtained via a cross-correlation with the Green Bank 4.85 GHz
catalogue (GB6, Gregory et al. 1996) covering 6.07 sr ($48.3\%$ of
the sky) to a flux limit of $S_{4.85}=18\,$mJy and comprising
$75,162$ sources\footnote{The cross-correlation was made using the
DIRA2 database and software (Nanni \& Tinarelli 1993; Battistini
et al. 1994) developed by the Astronet Database Group -- Italy,
available at the Institute of Radio Astronomy (IRA) of the
National Research Council (CNR). DIRA2 is at
http://www.ira.bo.cnr.it/dira2/}. The distribution of the
polarization degree $\Pi = (U^2+Q^2)^{1/2}/I$ at 1.4 GHz for a
complete sub-sample of  NVSS sources brighter than
$S(1.4\,\hbox{GHz})\ge 80\,$mJy is shown in
Fig.~\ref{poldistrflatsteep} for steep- ($\alpha > 0.5$) and
flat/inverted-spectrum ($\alpha \le 0.5$) sources separately. Note
that the the low--$\Pi$ tail of the distribution is contaminated
by the residual instrumental polarization which is estimated to be
$\sim 0.3\%$ (Condon et al. 1998). For this reason, in
investigating correlations, sources for which the NVSS catalog
yields a polarization degree $\Pi \le 1\%$ are attributed an upper
limit of 1\%.


As illustrated by Figs.~\ref{poldegvsfluxs} and
\ref{poldegvsfluxf}, the mean polarization degree is
anti-correlated with the flux density, especially in the case of
steep-spectrum sources. The median $\Pi$ steadily decreases from
$1.8\%$ for the 100--200 mJy bin (both for steep- and
flat-spectrum sources) to $<1\%$, for steep-spectrum sources, or
to $1.05\%$ for flat-spectrum sources, at flux densities $>
800\,$mJy.

We have tested the statistical significance of the correlation
using the computer package ASURV Rev 1.2, developed by Isobe, La
Valley \& Feigelson (La Valley et al. 1992), which implements the
methods presented in Feigelson \& Nelson (1985) and
in Isobe et al. (1986). 
For sources with $S(1.4\hbox{GHz}) \ge 80\,$mJy the test by Cox
proportional hazard model yields a global $\chi^2$,  with one
degree of freedom, of 573 for flat-spectrum and of 800 for
steep-spectrum sources. In both cases the null hypothesis is
rejected to a very high level of significance (probability $\ll
10^{-5}$).

The origin of this correlation is unclear.  One possibility is
that it may come from a change in the composition of the source
population with decreasing flux density. In fact, Snellen et al.
(2002) find that the fraction of flat-spectrum radio sources
identified with point sources (i.e. quasars) in APM scans
decreases, compared to the fraction of extended objects (i.e.
galaxies), with decreasing radio flux density. But the
distributions (see Fig.~\ref{distrpolsnellen}) of the polarization
degrees of NVSS compact and extended sources with
$S(1.4\hbox{GHz}) \ge 500\,$mJy in the sample by Snellen et al.
(2002) show that extended sources tend to be less polarized than
the compact ones, and therefore their effect goes in the direction
opposite to the observed anti-correlation. We caution however that
the situation may be different for steep-spectrum sources.

As a second attempt, we exploited the redshift data determined by
Sadler et al. (2002) by cross-matching the NVSS with the first 210
fields observed in the 2dF Galaxy Redshift Survey, covering an
effective area of 325 square degrees. There is an indication that
the sources with the highest radio power tend to be less polarized
at 1.4 GHz (see Fig.~\ref{polvslumEdS}). The standard statistical
tests, taking into account upper limits, do not detect, however, a
significant correlation between polarization degree and
luminosity. On the other hand, as will be seen in the next
section, there are clear evidences of substantial Faraday
depolarization at 1.4 GHz, at least for steep-spectrum sources.
But if the {\it intrinsic} polarization is uncorrelated with radio
luminosity, we would expect a positive correlation between {\it
observed} polarization and luminosity, since the more luminous
sources are, on average, at higher redshifts and the Faraday
rotation measures (RM) are proportional to $(1+z)^{-2}$. Thus, the
null result may be indicative of an anti-correlation between the
{\it intrinsic} polarization degree and radio luminosity or of an
increase of the intrinsic RM with $z$ compensating the decrease
due to the cosmological shift of frequencies to the red. Evidences
in favour of the latter possibility have been reported by
Pentericci et al. (2000). In the case of galaxies with intense
star-formation activity we may expect higher depolarization at
higher luminosities as the effect of both more chaotic magnetic
fields and of higher RMs, due to a higher abundance of ionized gas
and stronger magnetic fields.

If, on the other hand, the polarization degree is uncorrelated
with luminosity, the observed anti-correlation of the polarization
degree with flux density might be interpreted in terms of a
decreasing effect of Faraday depolarization if the average
redshift of sources increases with decreasing flux density.

\section{Faraday depolarization}

The exploitation of the NVSS data to estimate the contamination of
CMB polarization maps by radio sources requires the extrapolation
of 1.4 GHz polarized fluxes to high frequencies ($\nu \geq
20\,$GHz) as used for CMB polarization experiments. At the latter
frequencies the Faraday depolarization is probably small or
negligible, while it is generally important at 1.4 GHz.

An extensive multifrequency study of the linear polarization
properties of a representative sub-sample, comprising 192 sources,
of the B3 VLA survey (Vigotti et al. 1989) has been recently
carried out by Mack et al. (2002), using the Effelsberg 100-m
telescope. They made polarization measurements at 2.695 and 4.85
GHz of sources they had earlier detected in polarization at 10.6
GHz. All sources in this sample were observed with a resolution
larger than their angular extension, which was found to be always
significantly smaller than $60''$, while the Effelsberg
beam-widths are $69''$ at 10.6 GHz, $143''$ at 4.85 GHz and
$261''$ at 2.7 GHz; this excludes any frequency-dependent
differential depolarization. Their data, combined with the NVSS
data at 1.4 GHz, have allowed them to estimate the rotation
measures for 143 sources.

\begin{figure*}
   \centering
\includegraphics[width=7in,height=4.5in]{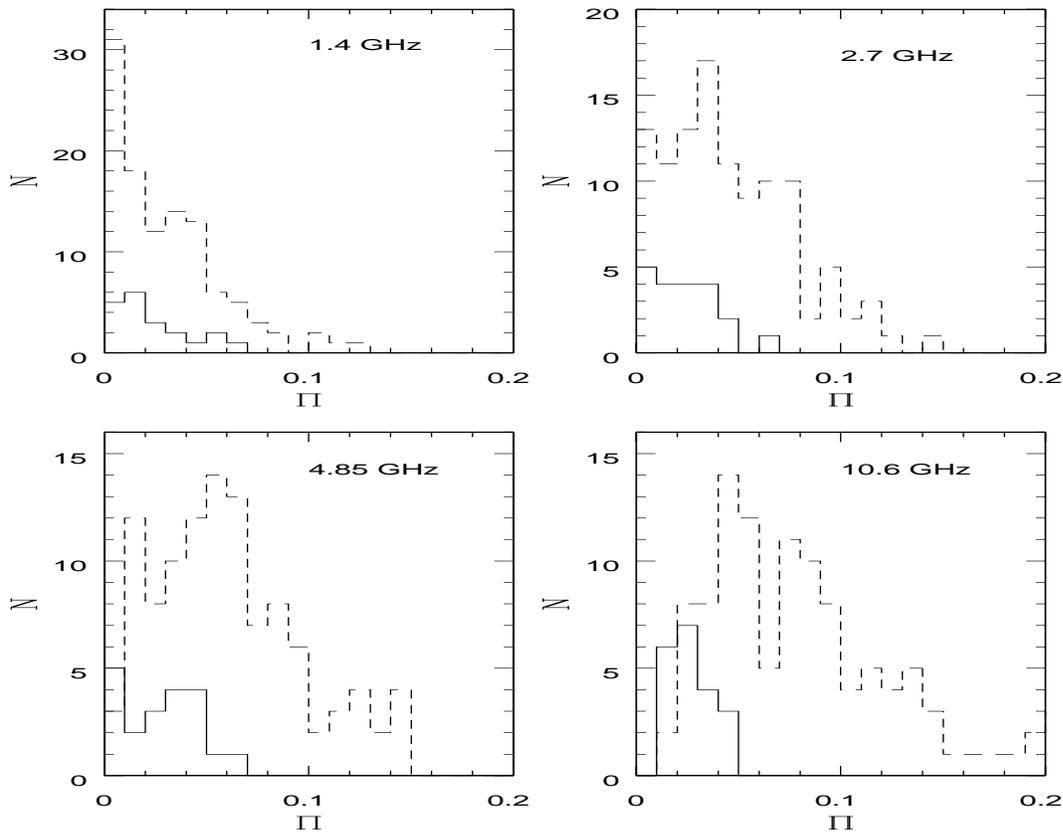}
      \caption{Polarization degree distributions of sources flat-
      (solid) and steep-spectrum (dashed) sources in the sample by
      Mack et al. (2002) at 4 frequencies.
              }
         \label{distrpolvigotti}
   \end{figure*}

As illustrated by Fig.~\ref{distrpolvigotti}, the distribution of
polarization degrees of steep-spectrum sources shifts to
increasingly higher values as the frequency increases, as expected
in the presence of substantial Faraday depolarization at 1.4 GHz.
For such sources we have $\langle\Pi\rangle= 2.93\%$, 4.68\%,
6.04\%, and 8.65\% at 1.4, 2.7, 4.85, and 10.6 GHz, respectively.
This frequency dependence of $\langle\Pi\rangle$ is consistent
with depolarization in uniform slabs with effective rotation
measure $\hbox{RM}\simeq 260$, implying that the Faraday
depolarization is small at $\ge 10\,$GHz. The mean polarization
degree at 1.4 GHz of these sources is significantly higher than
found for the full NVSS sample in the same flux density interval,
reflecting the selection criterion requiring sources to have
detected polarization at 10 GHz. On the other hand, since there is
no significant correlation between the ratio
$\Pi(10.4\,\hbox{GHz})/\Pi(1.4\,\hbox{GHz})$ and flux density at
either frequency, we assume that the mean ratio between the
polarization degrees at the two frequencies,
$\langle\Pi(10.4\,\hbox{GHz})/\Pi(1.4\,\hbox{GHz})\rangle \simeq
3$, is representative of the mean correction for Faraday
depolarization.

\begin{figure}
   \centering
\includegraphics[width=3in,height=3in]{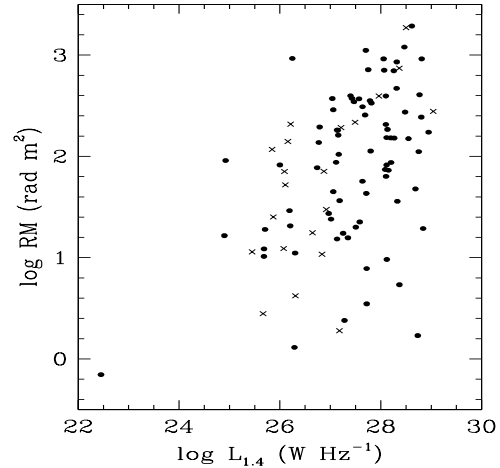}
      \caption{Rotation measures as a function of the 1.4 GHz luminosity
      for steep-spectrum sources in the sample by
      Mack et al. (2002) with spectroscopic (filled circles) or
      photometric redshift (crosses). An Einstein-de Sitter universe with
      $H_0=50$ has been adopted.
              }
         \label{RMvsLumVigotti}
   \end{figure}

The standard statistical tests applied to the 99 sources with
measured or estimated redshift detect a highly significant
correlation between {\it intrinsic} RMs (i.e. observed RM $\times
(1+z)^2$) and radio luminosity (see Fig.~\ref{RMvsLumVigotti}):
Cox's proportional hazard model gives $\chi^2 = 29$ for 1 degree
of freedom; Kendall's and Spearman's correlation tests give a
z-value of 4.5 and $\rho = 0.44$, respectively. According to each
test the null hypothesis (no correlation) has a probability $\ll
10^{-5}$. The correlation remains highly significant ($\chi^2 =
16$, probability of no correlation $\simeq 10^{-4}$) even if we
remove the point in the lower left-hand corner of
Fig.~\ref{RMvsLumVigotti}.


In the case of flat-spectrum sources we have, for the same
frequencies, $\langle\Pi\rangle= 1.66\%$, 2.19\%, 2.63\%, and
2.66\%, respectively, consistent with the Faraday depolarization
being relatively small, for most sources of this type, already at
1.4 GHz. This result is in agreement with earlier multifrequency
polarization studies of compact flat-spectrum sources (Jones et
al. 1985; Rudnick et al. 1985), which found a median value of
polarization $\sim 2.5\%$ between 1.4 and 90 GHz, independent of
frequency, although some sources do show a systematic increase of
polarization with frequency (see Fig.~\ref{distrpolvigotti} and
Rudnick et al. 1978). It should be noted, however, that Mack et
al. (2002) could not determine reliable RMs for flat-spectrum
sources. On the other hand, the mm/sub-mm polarization survey of
flat-spectrum radio sources by Nartallo et al. (1998) has yielded
significantly higher polarization degrees. If for each object we
take the mean of measurements made at different epochs, we find
that the median polarization degree at $1.1\,$mm of High
Polarization Quasars (10 objects) is 7.3\%, that of Low
Polarization Quasars (4 objects) is 4.3\%, and that of BL Lacs (11
objects) is 6.3\%. For comparison, the median polarization degrees
given by the NVSS for the same objects are 2.4\% for HPQs, $< 1\%$
for LPQs, and 1.9\% for BL Lacs. Only the 3 flat-spectrum
radio-galaxies in the sample by Nartallo et al. (1998) have a
polarization degree at $1.1\,$mm comparable or lower than that
measured by the NVSS. The median $1.1\,$mm polarization for the
full sample by Nartallo et al. (1998) is 5.5\%; at 1.4 GHz its
median polarization is 2\%. This would again suggest a factor of 3
increase of the polarization degree from 1.4 GHz to high
frequencies.


\section{Power spectrum estimates}

As mentioned above, we aim at estimating the power spectrum of
polarization fluctuations due to extragalactic radio sources at
the high frequencies ($\gsim 20\,$GHz) at which CMB polarization
experiments are carried out. We will make reference, in
particular, to the frequencies of the Low Frequency Instrument
(LFI) of ESA's {\sc Planck} mission (30, 44, 70, and 100 GHz),
which span a range similar to that covered by NASA's Microwave
Anisotropy Probe (MAP) mission (22, 30, 40, 60, 90 GHz).

Based on the discussion in the previous section, we assume that
the polarization degree of steep-spectrum sources increases, on
average, by a factor of 3 from 1.4 to 10 GHz and stays constant at
still higher frequencies. As for flat-spectrum sources, we
consider two possibilities: their polarization degree is either
frequency-independent, or increases, on average, by a factor of 3
at high frequencies, as is the case for sources in the sample by
Nartallo et al. (1998). In the latter case, we assume that the
increase applies to frequencies $\nu \geq 30\,$GHz.

A second issue is the extrapolation of source spectra. Adoption of
the 1.4--4.85 spectral indices up to high frequencies leads to
over-predicting the surface density of inverted- ($\alpha<0$),
flat- ($0\le \alpha \le 0.5$), and steep-spectrum sources brighter
than $30\,$mJy at $15\,$GHz by a factor of 9.2, 2.2, and 1.2,
respectively, compared to the findings by Taylor et al. (2001).
This indicates average steepenings above 4.8 GHz of
$\Delta\alpha_{\rm steep} \simeq 0.12$, $\Delta\alpha_{\rm flat}
\simeq 0.47$, and $\Delta\alpha_{\rm inverted} \simeq 1.35$, that
we incorporate in our flux density extrapolations.

In estimating the power spectra we assume that sources above the
detection limits estimated by Toffolatti et al. (1998)  for the
{\sc Planck} channels, namely 650, 480, 330, and 350 mJy at 30,
44, 70, and 100 GHz, respectively, are removed.

We have selected the portion of the sky region covered by both the
NVSS and the GB6 survey at $b>10^\circ$ (area of 16,726 square
degrees). The power spectrum has been computed using the HEALPix
package (G\'orski et al. 1999). An analysis of the NVSS data alone
(which reach lower fluxes and therefore allow us to derive the
power spectrum for a broader range of angular scales) shows no
evidence for departures from Poisson noise (which produces a
simple white noise power spectrum, with the same power in all
multipoles). The same conclusion was reached by Tegmark \&
Efstathiou (1996) from an analysis of a point-source catalogue
(Becker et al. 1995) from the VLA FIRST survey. This is not in
conflict with the evidences for clustering of radio-sources in the
GB6 (Kooiman et al. 1995) and in the FIRST (Cress et al. 1996)
survey. In fact, as shown by Toffolatti et al. (1999), the
contribution of the observed clustering to the power spectrum of
the source distribution is small in comparison with the Poisson
contribution, not surprisingly in view of strong dilution of the
clustering signal, due to the broad redshift distribution of radio
sources. The same conclusion holds for steep- and flat- plus
inverted-spectrum sources separately, although in this case the
range of scales that can be investigated is more limited, due to
the higher flux limit of the GB6 survey.

The power spectrum of fluctuations due to a Poisson distribution
of sources whose differential source count per steradian, as a
function of the flux density $S$, is $n(S)$, writes (Tegmark \&
Efstathiou 1996):
\begin{equation}
C_\ell = \int_0^{S_c} n(S)\, S^2\, dS\ ,
\end{equation}
where $S_c$ is the minimum flux density of sources that can be
individually detected and removed. Clearly we can derive only a
lower limit to $C_\ell$ because we can estimate the high frequency
counts only for sources brighter than $S_{4.85\,{\rm GHz}} \ge
18\,$mJy, and we are therefore missing the contribution of fainter
sources. However, since the slope of $n(S)$ in the relevant flux
density range is $\beta = -d\log n/d\log S \simeq 2$ (Taylor et
al. 2001), the underestimate of $C_\ell$ is probably no more than
$\sim 10\%$. We have tested this by attributing to NVSS sources
not detected at 4.85 GHz a spectral index drawn at random from the
spectral index distribution of sources also present in GB6
catalogue, after having checked that no dependence of such a
distribution on flux density is indicated by the (very limited)
data on sources down to $S(1.4\,\hbox{GHz})\gsim 1\,$mJy (Fomalont
et al. 1984; Donnelly et al. 1987). This implies extending the
high-frequency counts of "flat''-spectrum sources downwards in
flux by a factor $\simeq 7$ (and by a factor of $\simeq 20$ those
of the steep-spectrum ones, which however yield a smaller
contribution to fluctuations at high frequencies). As expected,
the $C_\ell$'s increased by only a few percent. We also
investigated the effect of missing polarization measurements by
attributing to sources whose polarization was not measured a
polarization degree and a polarization angle drawn at random from
the observed distributions. Again, the derived power spectra did
not change appreciably.

As expected (Seljak 1997), sources yield essentially identical
contributions to $E$- and $B$-mode power spectra; therefore only
the $E$-mode is plotted in Fig.~\ref{Cl}. We have also checked
that the $TE$ ($T$ representing temperature fluctuations) power
spectrum vanishes, as expected due to the random distribution of
polarization position angles of sources.


\section{Discussion and conclusions}

Until adequate high frequency ``blind'' polarization surveys will
become available, estimates of the contamination of CMB
polarization maps by extragalactic sources will require delicate
extrapolations of data from low-frequency surveys.

The main strength of the present analysis, compared with previous
ones, is the use of a far more extended data-base. The NVSS survey
has yielded polarization data for a very large, complete sample of
extragalactic sources, allowing a direct observational
determination of the power spectrum of polarization fluctuations
due to them at $1.4\,$GHz. The data also show a previously
unnoticed anti-correlation of the polarization degree with flux
density, which deserves further investigation.

Furthermore, coupling NVSS with GB6 data made possible to
separately determine the contributions of steep and
``flat''-spectrum sources. The median polarization degree of the
two populations for $S(1.4\,\hbox{GHz}) \ge 80\,$mJy turns out to
be essentially equal (at the level of 2.2\%) and shows a similar
decrease with increasing flux density.

The data by Mack et al. (2002) have allowed us to investigate how
the polarization properties of sources vary with frequency. For
steep-spectrum sources we found clear evidence of Faraday
depolarization corresponding to an effective value of the rotation
measure $\hbox{RM} \simeq 260 \hbox{rad}\,\hbox{m}^{-2}$. The
average intrinsic polarization degree of these sources, observed
at $\nu \gsim 10\,$GHz, is estimated to be, on average, 3 times
higher than at 1.4 GHz.

On the other hand, the mean polarization degree of
flat-/inverted-spectrum sources increases only weakly from 1.4 to
10.6 GHz, suggesting that these sources have either small or
really extreme RMs. Of course, polarization properties of radio
sources may change with frequency not only by Faraday rotation but
also for other reasons, such as the appearance of new emission
components with different polarization properties. Only
high-frequency polarization surveys may resolve this issue. We
have considered two possibilities: either the mean polarization
degree is frequency-independent, as indicated by the
multifrequency studies by Jones et al. (1985) and Rudnick et al.
(1985), or it increases, at high frequencies, by a factor $\simeq
3$ compared to 1.4 GHz, as is the case for the sample of Nartallo
et al. (1998).

Another critical issue is the extrapolation in frequency of the
observed flux densities. Adoption of the observed 1.4--4.85 GHz
spectral indices leads to over-predicting the 15 GHz counts by a
factor $\simeq 3$, compared with results of the survey by Taylor
et al. (2001). Predictions of the most commonly used evolutionary
models (Dunlop \& Peacock 1990; Toffolatti et al. 1998),
accounting for existing source counts up to 8.4 GHz and for the
associated redshift/luminosity distributions, yield 15 GHz counts
in excess by similar factors. Therefore, to extrapolate the 1.4
GHz flux densities we have exploited the observed spectral indices
only up to 5 GHz. Above this frequency we have adopted the average
spectral steepenings necessary to ensure, for each population
(steep-, flat- and inverted-spectrum sources), consistency with
the results by Taylor et al. (2001).

%
%
  \begin{figure*}
   \centering
\includegraphics[width=7in,height=5in]{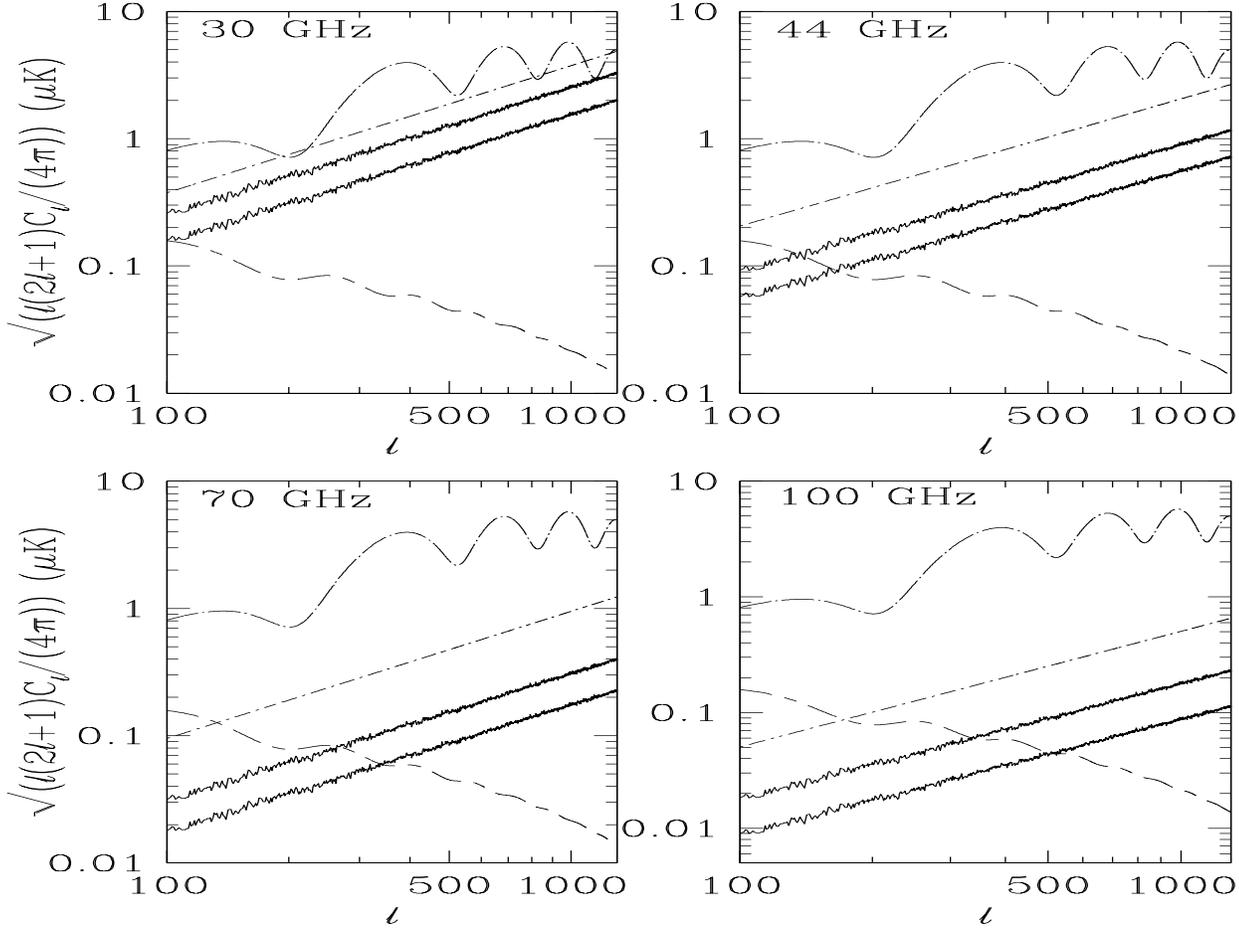}
\caption{Power spectrum of polarization fluctuations due to
extragalactic radio sources (irregular lines; $E$- and $B$-modes
are indistinguishable) at {\sc Planck}-LFI frequencies: 30 GHz
(upper left-hand panel), 44 GHz (upper right-hand panel), 70 GHz
(lower right-hand panel), and 100 GHz (lower right-hand panel).
The lower irregular lines correspond to the case of a
frequency-independent polarization degree for flat- and inverted
spectrum sources, while the upper ones correspond to a factor of 3
increase of the mean polarization degree at Planck frequencies,
compared to NVSS measurements. In both cases the mean polarization
degree of steep-spectrum sources measured by the NVSS survey has
been corrected upwards by a factor of 3. The dot-dashed straight
line shows the preliminary estimate by De Zotti et al. (1999).
Also shown, for comparison, are the CMB $E$- (dot-dashed curve)
and $B$-mode (long-short dashes) power spectra for a flat CDM
cosmological model with $\Omega_\Lambda =0.7$, $\Omega_{\rm dark
matter}=0.25$, $\Omega_{\rm baryon}=0.05$,
$H_0=70\,\hbox{km}\,\hbox{s}^{-1}\,\hbox{Mpc}^{-1}$, and a tensor
contribution to the temperature quadrupole equal to 30\% of that
of scalar perturbations. The CMB power spectra were computed with
CMBFAST (Seljak \& Zaldarriaga 1996).
              }
         \label{Cl}
   \end{figure*}

An estimate of the power spectrum of polarization fluctuations at
{\sc Planck} frequencies obtained extrapolating the $1.4\,$GHz
fluxes as described in Sect.$\,$2, with an upper flux-density
cut-off corresponding to the source detection limit in each {\sc
Planck} channel as estimated by Toffolatti et al. (1998) and
applying the average correction for Faraday depolarization derived
in Sect.$\,$3, is shown in Fig.~\ref{Cl}. The present estimates
are significantly below those by De Zotti et al. (1999), mostly
due to the substantial steepening of the source spectra between 5
and $15\,$GHz implied by the results of the survey by Taylor et
al. (2001). As illustrated by Fig.~\ref{Cl}, extragalactic radio
sources are not expected to be a serious hindrance for
measurements of the CMB $E$-mode power spectrum at $\nu \geq
30\,$GHz. They are even less of a problem for measurements of the
temperature-$E$-mode correlation since their $TE$ power spectrum
vanishes, owing to the random distribution of their polarization
position angles.

\begin{acknowledgements}

We thank E. Sadler and I. Snellen for having made available their
data in computer readable form, and M. Nanni for assistance in the
use of the DIRA2 database.
KHM was supported through a European Community Marie-Curie fellowship.
Work supported in part by ASI and MIUR.

\end{acknowledgements}


\begin{thebibliography}{}










\bibitem[]{}


\bibitem[1994]{battistini}
{Battistini}, P., {Benacchio}, L., {Claudi}, R.U., \& {Sarasso},
M. 1994, {DIRA2 Database: The Catalogue Documentation}, (Astronet
Special Publication, Osservatorio Astrofisico di Arcetri, Firenze)

\bibitem[Becker et al.~1995]{1995ApJ...450..559B} Becker, R.H., White
R.L., \& Helfand D.J. 1995, ApJ,  450, 559

\bibitem[]{} Cecchini, S., Cortiglioni, S., Sault, R., \& Sbarra, C.
(eds.) 2002, Astrophysical Polarized Backgrounds, AIP Conf. Proc.
609


\bibitem[]{} Condon, J.J., Cotton, W.D., Greisen, E.W., et al. 1998 AJ,  115, 1693

\bibitem[Cress et al.~1996]{1996ApJ...473....7C} Cress, C.M.,
Helfand, D.J., Becker, R.H., Gregg, M.D., \& White, R.L. 1996,
ApJ, 473, 7

\bibitem[de Bernardis et al.~2002]{2002ApJ...564..559D} de Bernardis, P.,
Ade, P.A.R., Bock, J.J., et al. 2002, ApJ,  564, 559

\bibitem[]{} De Zotti, G. 2002, AIP Conf. Ser. 609, Astrophysical
Polarized Backgrounds, eds. S. Cecchini, S. Cortiglioni, R. Sault,
and C. Sbarra eds., 295

\bibitem[]{} De Zotti, G., Gruppioni, C., Ciliegi, P., Burigana,
C., \& Danese, L. 1999, New Astr., 4, 481

\bibitem[Donnelly et al.~1987]{1987ApJ...321...94D} Donnelly, R.H.,
Partridge, R.B., \& Windhorst, R.A. 1987, ApJ,  321, 94

\bibitem[Dunlop \& Peacock 1990]{1990MNRAS.247...19D} Dunlop, J.S.,
\& Peacock, J.A. 1990, MNRAS,  247, 19

\bibitem[]{} Feigelson E.D., \& Nelson P.I. 1985, ApJ,  293,
192

\bibitem[Fomalont et al.~1984]{1984Sci...225...23F} Fomalont E.B.,
Kellermann K.I., Wall J.V., \& Weistrop D. 1984, Science,  225, 23

\bibitem[G\'{o}rski et al. 1999]{GORSKI} G\'{o}rski K.M., Wandelt B.D.,
Hansen F.K.,  Hivon E., \& Banday A.J. 1999, astro-ph/9905275

\bibitem[]{} Gregory P.C., Scott W.K., Douglas K., \& Condon J.J.
1996, ApJS,  103, 427

\bibitem[Halverson et al.~2002]{2002ApJ...568...38H} Halverson, N.W.,
Leitch, E.M., Pryke, C., et al. 2002, ApJ,  568, 38

\bibitem[]{} Hedman, M.M., Barkats, D., Gundersen, J.O., Staggs, S.T.,
\& Winstein, B. 2001, ApJ, 548, L111

\bibitem[]{} Hu, W., \& Dodelson, S. 2002, ARA\&A, in press

\bibitem[]{} {Hu}, W., \& {White}, M. 1997, New Astr., 2, 323

\bibitem[]{} Isobe, T., Feigelson, E.D., \& Nelson, P.I., 1986, ApJ,
306, 490


\bibitem[]{} Jones, T.W., Rudnick, L., Fiedler, R.L., et al. 1985, ApJ,  290, 627

\bibitem[]{} Kamionkowski, M., \& Kosowsky, A. 1999, Ann. Rev. Nucl. Part.
Sci., 49, 77

\bibitem[]{Kamionkowski1997} Kamionkowski, M., Kosowsky, A., \& Stebbins, A. 1997, Phys. Rev.
D, 55, 7368

\bibitem[]{} Keating, B.G., O'Dell, C.W., de Oliveira-Costa, A., et
al. 2001, ApJ, 560, L1

\bibitem[Kooiman et al.~1995]{1995ApJ...448..500K} Kooiman, B.L.,
Burns, J.O., \& Klypin, A.A. 1995, ApJ, 448, 500

\bibitem[Lavalley et al.~1992]{1992adass...1..245L} LaValley, M., Isobe, \& T.,
Feigelson, E. 1992, ASP Conf.~Ser.~25: Astronomical Data Analysis
Software and Systems I,  1, 245

\bibitem[Lee et al.~2001]{2001ApJ...561L...1L} Lee, A.T., Ade, P., Balbi, A.,
et al. 2001, ApJ,  561, L1

\bibitem[]{} Mack, K.-H., Vigotti, M., Gregorini, L., \& Klein, U.
2002, in preparation

\bibitem[Miller et al.~1999]{1999ApJ...524L...1M} Miller, A.D.,
Caldwell, R., Devlin, M.J., et al. 1999, ApJ,  524, L1

\bibitem[1993]{nanni}
{Nanni}, M., \& {Tinarelli}, F. 1993, Mem. SAIt 64, 1053

\bibitem[]{} Nartallo, R., Gear, W.K., Murray, A.G., Robson, E.I., \& Hough, J.H. 1998,
MNRAS, 297, 667

\bibitem[]{} Pearson, T.J., Mason, B.S., Readhead, A.C.S., et al.
2002, AJ, submitted (astro-ph/0205388)

\bibitem[Pentericci et al.~2000]{2000A&AS..145..121P} Pentericci, L., Van
Reeven, W., Carilli, C.L., R{\"o}ttgering, H.J.A., \& Miley G.K.
2000, A\&AS,  145, 121

\bibitem[]{} Rudnick, L., Jones, T.W., Fiedler, R.L., et al. 1985, ApJS,  57,
693

\bibitem[]{} Rudnick, L., Owen, F.N., Jones, T.W., Puschell, J.J., \& Stein, W.A.
1978, ApJ,  225, L5

\bibitem[]{} Sadler E.M., Jackson C.A., Cannon R.D., et al. 2002,
MNRAS,  329, 227

\bibitem[]{} Sazhin M.V., \& Korol\"ev V.A. 1985, Pis'ma Astr. Zh., 11,
490, 1985 [Sov. Astr. Lett., 11, 204, 1985]

\bibitem[Seljak 1997]{1997ApJ...482....6S} Seljak, U. 1997, ApJ,  482, 6

\bibitem[]{} Snellen, I.A.G., McMahon, R.G., Hook, I.M., \& Browne, I.W.A.
2002, MNRAS,  329, 700

\bibitem[]{} Staggs, S.T., Gundersen, J.O., \& Church, S.E. 1999, in
ASP Conf. Ser. 181, Microwave Foregrounds, eds. A. de
Oliveira-Costa \& M. Tegmark, 299


\bibitem[]{} Taylor, A.C., Grainge, K., Jones, M.E., et al. 2001, MNRAS, 327, L1

\bibitem[Tegmark \& Efstathiou 1996]{1996MNRAS.281.1297T} Tegmark, M.,
\& Efstathiou, G. 1996, MNRAS,  281, 1297


\bibitem[]{} Toffolatti, L., Arg\"ueso-Gomez, F., De Zotti,
G., et al. 1998, MNRAS, 297, 117

\bibitem[]{} Toffolatti, L., De Zotti, G., Arg\"ueso, F., \& Burigana, C.
1999, in ASP Conf. Ser. 181, Microwave Foregrounds, eds. A. de
Oliveira-Costa \& M. Tegmark, 153

\bibitem[]{} Vigotti, M., Grueff, G., Perley, R., Clark, B.G.,
\& Bridle, A.H. 1989, AJ,  98, 419

\bibitem[]{ZaldarriagaSeljak1997} Zaldarriaga, M., \& Seljak, U.
1997, Phys. Rev. D, 55, 1830

\end{thebibliography}
\end{document}